\documentclass[%
superscriptaddress,
amsmath,amssymb,
aps,
twocolumn,
]{revtex4-1}

\usepackage{hyperref}
\hypersetup{colorlinks,allcolors=blue}
\usepackage{graphicx}
\usepackage{bm}
\usepackage[all]{hypcap} 
\usepackage{color}
\usepackage{siunitx}

\begin{document}

\title{Orbital and Spin Character of Doped Carriers in Infinite-Layer Nickelates}

\author{M.~Rossi}
\affiliation{Stanford Institute for Materials and Energy Sciences, SLAC National Accelerator Laboratory, 2575 Sand Hill Road, Menlo Park, California 94025, USA}

\author{H.~Lu}
\affiliation{Stanford Institute for Materials and Energy Sciences, SLAC National Accelerator Laboratory, 2575 Sand Hill Road, Menlo Park, California 94025, USA}
\affiliation{Department of Physics, Stanford University, Stanford, California 94305, USA}

\author{A.~Nag}
\affiliation{Diamond Light Source, Harwell Campus, Didcot OX11 0DE, United Kingdom}

\author{D.~Li}
\affiliation{Stanford Institute for Materials and Energy Sciences, SLAC National Accelerator Laboratory, 2575 Sand Hill Road, Menlo Park, California 94025, USA}
\affiliation{Department of Applied Physics, Stanford University, Stanford, California 94305, USA}

\author{M.~Osada}
\affiliation{Stanford Institute for Materials and Energy Sciences, SLAC National Accelerator Laboratory, 2575 Sand Hill Road, Menlo Park, California 94025, USA}
\affiliation{Department of Materials Science and Engineering, Stanford University, Stanford, California 94305, USA}

\author{K.~Lee}
\affiliation{Stanford Institute for Materials and Energy Sciences, SLAC National Accelerator Laboratory, 2575 Sand Hill Road, Menlo Park, California 94025, USA}
\affiliation{Department of Physics, Stanford University, Stanford, California 94305, USA}

\author{B.~Y.~Wang}
\affiliation{Stanford Institute for Materials and Energy Sciences, SLAC National Accelerator Laboratory, 2575 Sand Hill Road, Menlo Park, California 94025, USA}
\affiliation{Department of Physics, Stanford University, Stanford, California 94305, USA}

\author{S.~Agrestini}
\affiliation{Diamond Light Source, Harwell Campus, Didcot OX11 0DE, United Kingdom}

\author{M.~Garcia-Fernandez}
\affiliation{Diamond Light Source, Harwell Campus, Didcot OX11 0DE, United Kingdom}

\author{Y.-D.~Chuang}
\affiliation{Advanced Light Source, Lawrence Berkeley National Laboratory, 1 Cyclotron Road, MS 6-2100, Berkeley, California 94720, USA}

\author{Z.~X.~Shen}
\affiliation{Stanford Institute for Materials and Energy Sciences, SLAC National Accelerator Laboratory, 2575 Sand Hill Road, Menlo Park, California 94025, USA}
\affiliation{Department of Physics, Stanford University, Stanford, California 94305, USA}
\affiliation{Geballe Laboratory for Advanced Materials, Stanford University, Stanford, California 94305, USA}

\author{H.~Y.~Hwang}
\affiliation{Stanford Institute for Materials and Energy Sciences, SLAC National Accelerator Laboratory, 2575 Sand Hill Road, Menlo Park, California 94025, USA}
\affiliation{Department of Applied Physics, Stanford University, Stanford, California 94305, USA}
\affiliation{Geballe Laboratory for Advanced Materials, Stanford University, Stanford, California 94305, USA}

\author{B.~Moritz}
\affiliation{Stanford Institute for Materials and Energy Sciences, SLAC National Accelerator Laboratory, 2575 Sand Hill Road, Menlo Park, California 94025, USA}

\author{Ke-Jin~Zhou}
\affiliation{Diamond Light Source, Harwell Campus, Didcot OX11 0DE, United Kingdom}

\author{T.~P.~Devereaux}
\email[Corresponding author: ]{tpd@stanford.edu}
\affiliation{Stanford Institute for Materials and Energy Sciences, SLAC National Accelerator Laboratory, 2575 Sand Hill Road, Menlo Park, California 94025, USA}
\affiliation{Department of Materials Science and Engineering, Stanford University, Stanford, California 94305, USA}
\affiliation{Geballe Laboratory for Advanced Materials, Stanford University, Stanford, California 94305, USA}

\author{W.~S.~Lee}
\email[Corresponding author: ]{leews@stanford.edu}
\affiliation{Stanford Institute for Materials and Energy Sciences, SLAC National Accelerator Laboratory, 2575 Sand Hill Road, Menlo Park, California 94025, USA}


\begin{abstract}
The recent discovery of superconductivity in Nd$_{1-x}$Sr$_{x}$NiO$_2$ has drawn significant attention in the field. A key open question regards the evolution of the electronic structure with respect to hole doping. Here, we exploit x-ray absorption spectroscopy (XAS) and resonant inelastic x-ray scattering (RIXS) to probe the doping dependent electronic structure of the NiO$_2$ planes. Upon doping, a higher energy feature in Ni $L_3$ edge XAS develops in addition to the main absorption peak. By comparing our data to atomic multiplet calculations including $D_{4h}$ crystal field, the doping induced feature is consistent with a $d^8$ spin singlet state, in which doped holes reside in the $d_{x^2-y^2}$ orbitals, similar to doped single band Hubbard models. This is further supported by orbital excitations observed in RIXS spectra, which soften upon doping, corroborating with Fermi level shift associated with increasing holes in the $d_{x^2-y^2}$ orbital.
\end{abstract}

\maketitle

Infinite-layer nickelates have been proposed more than two decades ago as promising materials that may host unconventional superconductivity \cite{Anisimov1999}. Inspired by the crystal and electronic structure of high-temperature superconducting cuprates, Anisimov \emph{et al.} \cite{Anisimov1999} investigated LaNiO$_2$ that shares some of their essential characteristics: two-dimensional NiO$_2$ planes, nominal $3d^9$ valence configuration, and a $3d_{x^2-y^2}$-derived band crossing the Fermi level. However, superconductivity in nickelates remained elusive until very recently, when Li \emph{et al.} reported the first experimental evidence of a zero-resistance state in thin films of Nd$_{0.8}$Sr$_{0.2}$NiO$_2$ \cite{Li2019} with a transition temperature around 10 K at optimal doping \cite{Li2020,Zeng2020}. 

Although the early theoretical investigation was motivated by the search for cuprate analogs \cite{Anisimov1999}, some differences between the parent compounds of cuprates and nickelates emerge \cite{Lee2004}. NdNiO$_2$ is a poor conductor \cite{Li2019} and lacks evidence of long-range magnetic order \cite{Hayward2003}. Recent x-ray absorption spectroscopy (XAS) and resonant inelastic x-ray scattering (RIXS) experiments \cite{Hepting2020} have indicated that the charge-transfer energy to the O ligands is large, resulting in a much reduced hybridization between Ni and O states with respect to cuprates and other charge-transfer compounds. The half-filled two-dimensional Ni $3d_{x^2-y^2}$ orbital behaves as a Hubbard-like band coupled to a three-dimensional Nd $5d_{3z^2-r^2}$ band that crosses the Fermi level. The coupling between Ni $3d$ and Nd $5d$ bands may resemble an Anderson or Kondo lattice typical of heavy fermion systems, where the role of magnetic ``impurities'' is played by strongly-correlated Ni $3d_{x^2-y^2}$ electrons \cite{Hepting2020}. 

Since superconductivity emerges in NdNiO$_2$ when trivalent Nd is substituted by divalent Sr \cite{Li2019,Li2020,Zeng2020}, it is important to establish the evolution of the electronic structure upon doping. To date, theories have proposed distinct scenarios that depend on the interplay between electronic correlation, charge-transfer energy, crystal field splitting and Hund’s coupling \cite{Hu2019,Jiang2020,Lechermann2020,Liu2020,Petocchi2020,Wan2020}. For example, doped holes may give rise to Ni $3d^8$ sites, where the two holes may arrange in a singlet (double occupation of $3d_{x^2-y^2}$ orbital) \cite{Anisimov1999,Kitatani2020,Zhang2020,Zhang2020a,Krishna2020} or triplet (occupation of different orbitals) \cite{Chang2019,Lechermann2020a,Werner2020,Zhang2020b} configuration; or they may form $d^9 \underline{L}$ states (where $\underline{L}$ denotes a hole in the O $2p$ orbitals) that resemble Zhang-Rice singlets \cite{Lang2020} typical of doped cuprates \cite{Zhang1988}; also, carriers may be introduced into the Nd $5d$ states \cite{Choi2020,Petocchi2020}. Experimental determination of the orbital and spin character of doped holes is therefore highly desirable to clarify this issue. 

Notably, using electron energy loss spectroscopy (EELS), Goodge \emph{et al.} \cite{Goodge2020} recently reported changes in the absorption spectra of Nd$_{1-x}$Sr$_x$NiO$_2$ at the O, Ni and Nd edges, suggesting that doped carriers appear to reside primarily on Ni sites forming $d^8$ states with little change in oxygen content -- much less than that observed in cuprates \cite{Chen1991}. Yet, further spectroscopic studies are needed to confirm these results and establish whether the two holes in the Ni $3d$ states form a singlet or triplet. 

In this Letter, we use a combination of high-resolution XAS and RIXS and multiplet calculations to investigate the orbital and spin states of doped holes in Nd$_{1-x}$Sr$_x$NiO$_2$. We consider three doping levels: $x = 0$ (undoped), 0.125 (non-superconducting), and 0.225 (superconducting, $T_c \approx 10$ K). XAS and RIXS are suitable techniques for our purpose since the resonant excitation grants element selectivity while enhancing the scattering cross-section, which is crucial for thin film samples. We find that doped holes are primarily introduced in the Ni $3d_{x^2-y^2}$ Hubbard band in a low-spin configuration associated with a shift of Fermi level that is reflected in the softening of orbital excitations by approximately 0.2 eV between compounds with Sr content of $x = 0$ and 0.225.

Thin films of the precursor perovskite Nd$_{1-x}$Sr$_x$NiO$_3$ with a thickness of 10 nm were grown on a substrate of SrTiO$_3$(001). The $c$-axis oriented infinite-layer Nd$_{1-x}$Sr$_x$NiO$_2$ was obtained by means of a topotactic reduction process \cite{Lee2020}. To protect and support the crystalline order, a capping layer made of five unit cells of SrTiO$_3$(001) was grown on top of the nickelate film. XAS and RIXS measurements were performed at beamline I21 of the Diamond Light Source (United Kingdom). The combined energy resolution of the beamline and the spectrometer was approximately 40 meV at the Ni $L_3$ edge. Measurements were taken at 20 K. RIXS spectra are normalized to the incident photon flux.

\begin{figure}
	\centering
	\includegraphics[width=0.7\columnwidth]{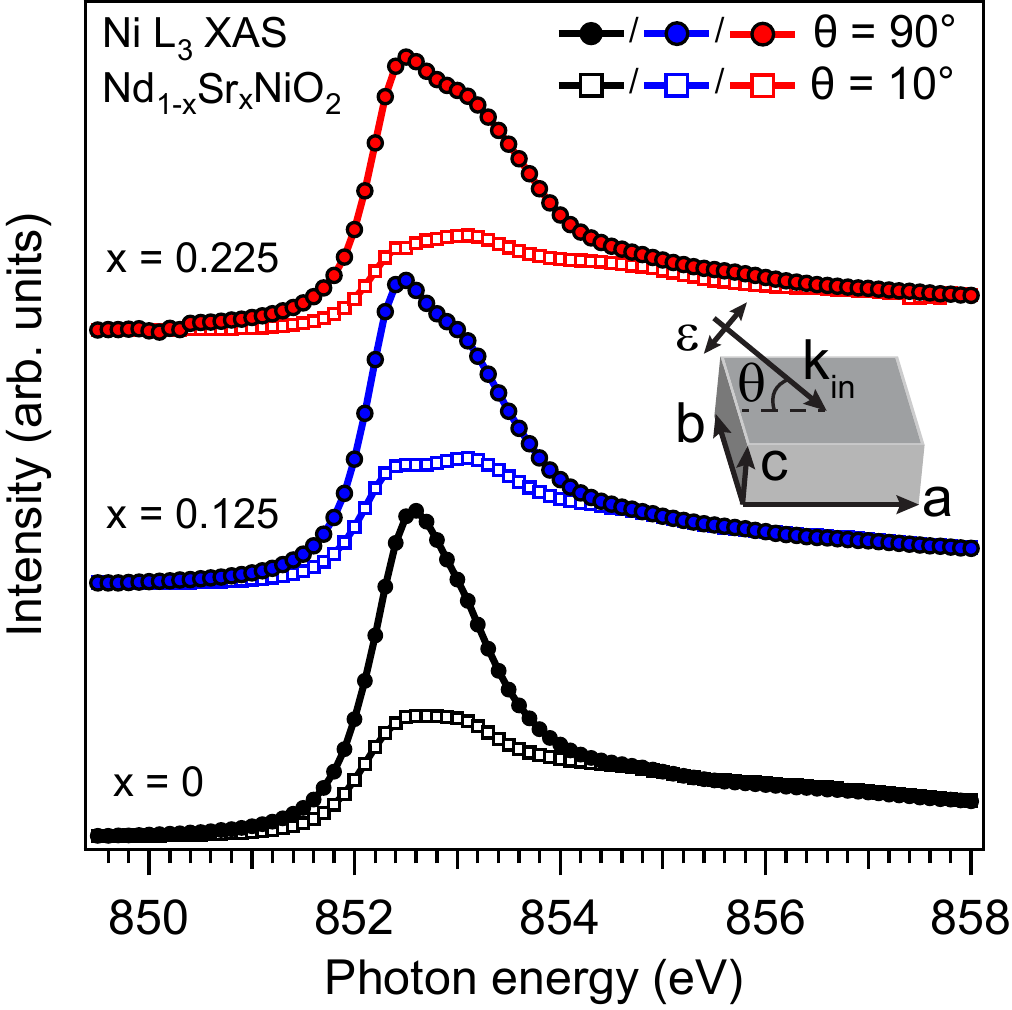}
	\caption{\label{fig1} Total electron yield Ni $L_3$ edge XAS of Nd$_{1-x}$Sr$_x$NiO$_2$. A constant background has been removed and the post-edge intensity has been set to 1. The drawing shows the direction of the photon polarization $\bm{\epsilon}$ in the sample reference frame $abc$.}
\end{figure}

\begin{figure}
	\centering
	\includegraphics[width=\columnwidth]{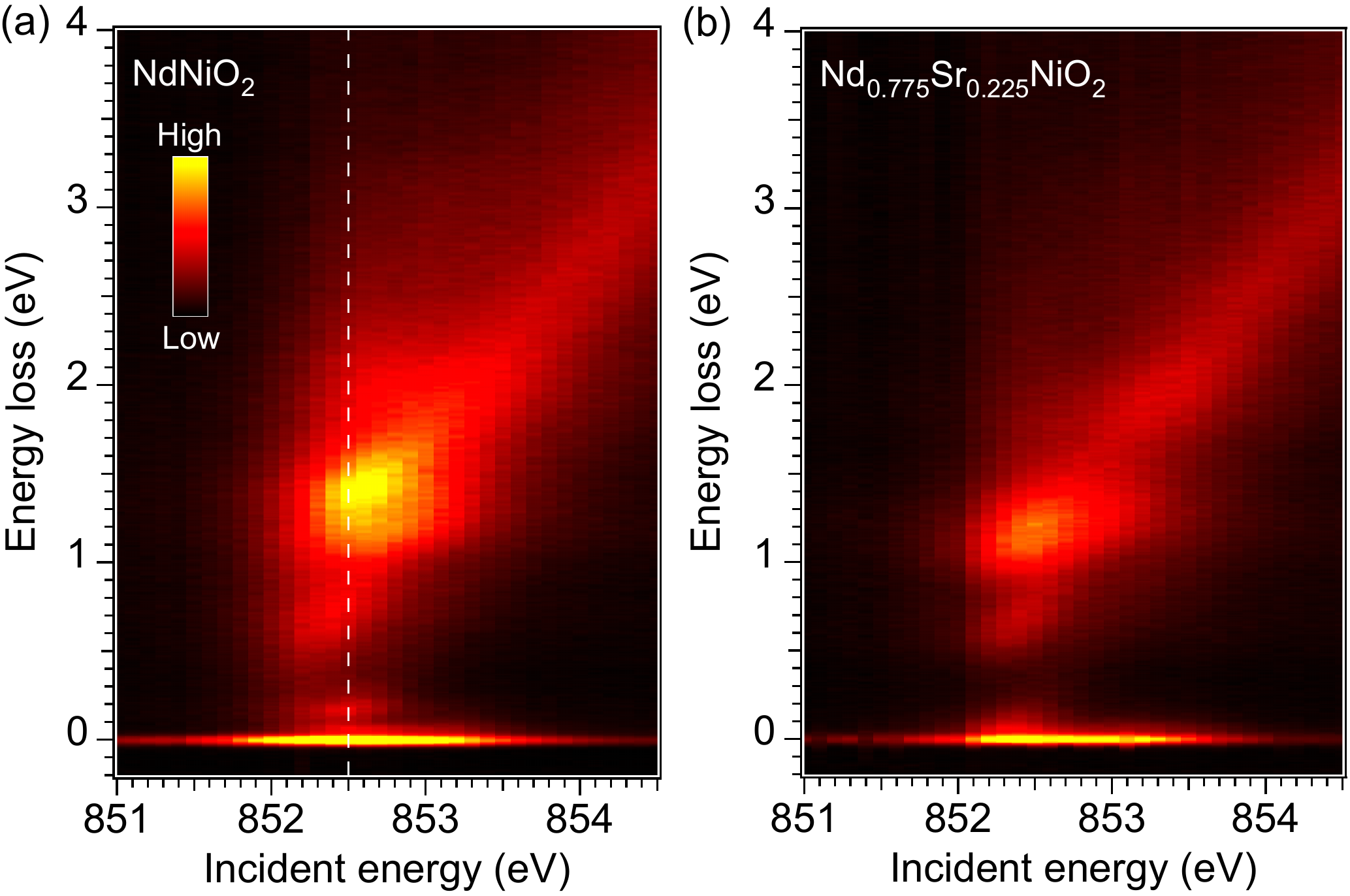}
	\caption{\label{fig2} RIXS intensity maps of NdNiO$_2$ (a) and Nd$_{0.775}$Sr$_{0.225}$NiO$_2$ (b) as a function of incident photon energy across the Ni $L_3$ edge. RIXS spectra were collected at an incidence angle of \SI{35}{\degree} and scattering angle of \SI{154}{\degree}. The dashed line marks the incident energy selected for the measurement of the angular dependence of RIXS spectra.}
\end{figure}

\begin{figure*}
	\centering
	\includegraphics[width=\textwidth]{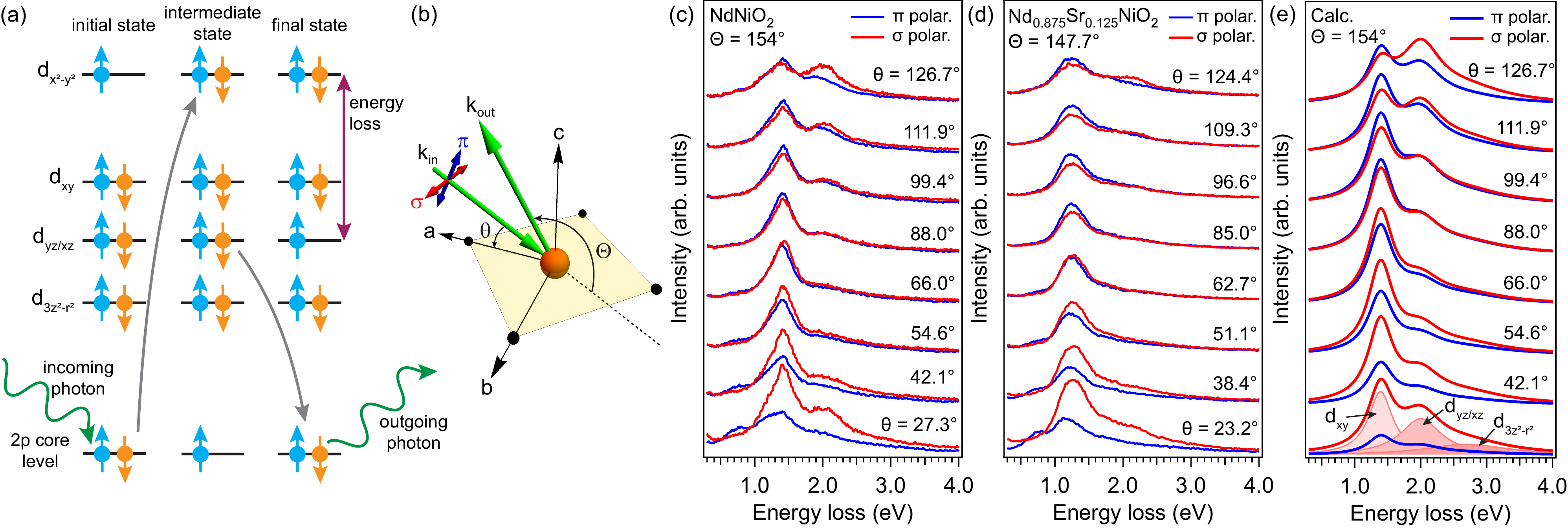}
	\caption{\label{fig3} (a) Schematics of the RIXS process at the Ni $L_3$ edge. (b) Ni cation (orange sphere) surrounded by O anions (black spheres) and sketch of the experimental geometry. The incoming photon wave vector $\mathbf{k}_\mathrm{in}$ forms an angle $\theta$ with the sample $ab$ plane. The scattered photon wave vector $\mathbf{k}_\mathrm{out}$ forms an angle $\Theta$ with $\mathbf{k}_\mathrm{in}$. The incident photon polarization is either parallel ($\pi$) or perpendicular ($\sigma$) to the scattering plane, which coincides with the $ac$ plane. (c, d) Stack plot of RIXS spectra of NdNiO$_2$ (c) and Nd$_{0.875}$Sr$_{0.125}$NiO$_2$ (d) collected at various incidence angles $\theta$ and fixed scattering angle $\Theta$. The incident energy was fixed to 852.5 eV for NdNiO$_2$ (dashed line in Fig.~\ref{fig2}(a)) and 852.44 eV for Nd$_{0.875}$Sr$_{0.125}$NiO$_2$. (e) Angular dependence of the intensity of $dd$ transitions calculated for the angles of panel (c).}
\end{figure*}

We first study the doping evolution of the electronic states of Nd$_{1-x}$Sr$_{x}$NiO$_2$ by examining XAS spectra at relevant Ni, O, and Nd absorption edges. Figure~\ref{fig1} shows Ni $L_3$ edge XAS spectra taken with two different light polarizations $\bm{\epsilon}$ set by the incidence angle $\theta$ (see inset for a sketch of the experimental geometry). The XAS spectrum of NdNiO$_2$ measured with $\theta = \SI{90}{\degree}$ ($\bm{\epsilon} \parallel a$) shows one dominant peak attributed to the $2p^63d^9$--$2p^53d^{10}$ transition \cite{Hepting2020,Goodge2020}. Upon doping, XAS spectra measured with $\bm{\epsilon} \parallel a$ develop an additional shoulder at higher energy than the main peak. This spectral feature is consistent with the broadening of the main absorption peak reported in EELS measurements \cite{Goodge2020}, but is unambiguously resolved in our high-resolution data. For the XAS taken at $\theta = \SI{10}{\degree}$ ($\bm{\epsilon}$ mostly along the $c$ axis), we observe a strong linear dichroism, suggesting a pronounced in-plane orientation of the unoccupied $3d$ orbitals.

XAS spectra of Nd$_{1-x}$Sr$_{x}$NiO$_2$ measured at the O $K$ edge (see Supplemental Material \cite{SM}) do not reveal the emergence of a pre-peak usually associated with hole doping into the charge-transfer band \cite{Chen1991}. While the signal from the SrTiO$_3$ capping layer could obscure the absorption from the underlying nickelate film, our results nonetheless suggest that the doping induced signature in O $K$ edge XAS is minor, as confirmed by the spatially-resolved EELS measurements \cite{Goodge2020}. The Nd $M_5$ edge XAS spectra are also unaffected by doping (see Supplemental Material \cite{SM}), implying well-localized Nd $4f$ states. Since calculations indicate that Nd $5d$ bands give rise to a metallic state \cite{Lee2004}, they should exhibit changes upon doping; however, spectroscopic validation of this point is beyond the scope of this work and will be addressed in future studies. Overall, our findings support the scenario in which the large charge-transfer energy favors the formation of Ni $d^8$ states rather than $d^9\underline{L}$ states.  This corroborates the classification of infinite-layer nickelates into the Mott-Hubbard region of the Zaanen-Sawatzky-Allen scheme \cite{Zaanen1985,Hepting2020,Goodge2020,Fu2019}.

After establishing that doped holes are introduced primarily into Ni $3d$ orbitals, we further analyze the energetics of $3d$ states and their doping dependence using RIXS. Figure~\ref{fig2} displays the RIXS intensity maps of NdNiO$_2$ (a) and Nd$_{0.775}$Sr$_{0.225}$NiO$_2$ (b) as a function of incident photon energy across the Ni $L_3$ edge. The most intense excitations are found at an energy loss of 1--3 eV, which is the typical energy range of crystal field ($dd$) transitions \cite{Ghiringhelli2004,Ghiringhelli2005,Moretti2011,Bisogni2016,Fabbris2017}. At an energy loss of $\approx 0.7$ eV, an excitation is visible in both compounds. This feature is specific to infinite-layer nickelates and has been attributed to the hybridization between Ni and Nd orbitals \cite{Hepting2020}. For incident photon energies above 853 eV, a fluorescence-like excitation is observed. Low energy-loss excitations below 0.3 eV will be discussed in separate works.

We focus on the $dd$ transitions that carry direct information about the Ni $3d$ orbitals. The energy and orbital character of the $dd$ excitations are analyzed using an ionic model including crystal field. Although simple, the model has proven successful to characterize the orbital excitations of cuprates \cite{Moretti2011}. We consider a single Ni$^{+}$ cation surrounded by four negative point charges in square planar geometry. The $D_{4h}$ tetragonal symmetry of the crystal field lifts the degeneracy of the Ni $3d$ orbitals such that the $d_{3z^2-r^2}$ becomes the lowest energy state, followed by $d_{yz/xz}$, $d_{xy}$ and $d_{x^2-y^2}$. In the $L_3$ edge RIXS process, illustrated in Fig.~\ref{fig3}(a), a $2p_{3/2}$ core electron is first promoted into the empty $3d_{x^2-y^2}$ orbital, then a $3d$ electron radiatively decays producing an electronic redistribution within the $3d$ shell that gives rise to \textit{dd} transitions in the RIXS spectra. Since the $3d$ orbitals have distinct spatial symmetry, their orientation relative to the polarization vectors of the incident and scattered photons produces a modulation of the RIXS matrix elements and therefore of the $dd$ intensity. Indeed, as shown in Figs.~\ref{fig3}(c,d), while the energy and width of the $dd$ excitations of a particular sample do not exhibit notable variation, the relative intensity varies as a function of the incidence angle $\theta$, which effectively alters the polarization direction relative to the orbital orientation (see the sketch of the experimental geometry in Fig.~\ref{fig3}(b)). This variation can be used to determine the orbital character of the $dd$ excitations and verify whether the $3d$ energy levels follow the expected $D_{4h}$ crystal field splitting.

We calculate the RIXS cross section of $dd$ transitions using the single ion model, following Ref.~\onlinecite{Moretti2011}, which is also described in the Supplemental Material \cite{SM}. We considered the incidence and scattering angles reported in Fig.~\ref{fig3}(c), but the calculations equally apply to panel (d) since the minor differences in the two experimental geometries do not appreciably affect the $dd$ intensity modulation. The cross sections, reported in panel (e), not only qualitatively reproduce the angular dependence of the experimental intensity for a given incident polarization, but also capture the relative modulation between $\pi$ and $\sigma$ polarizations at a given angle. This agreement allows us to attribute the peaks in the RIXS spectra of NdNiO$_2$ at 1.39 and 2.0 eV and the broad tail centered at 2.7 eV to transitions into the $d_{xy}$, $d_{yz/xz}$, and $d_{3z^2-r^2}$ orbitals, respectively, as shown in panel (e). A similar assignment can be made for Nd$_{0.875}$Sr$_{0.125}$NiO$_2$. We note that the weaker cross section of the $d_{3z^2-r^2}$ excitation and its larger width compared to other $dd$ transitions prevent us from precisely constrain its energy position. Nevertheless, our assignment is consistent with experimental and theoretical findings on CaCuO$_2$ \cite{Moretti2011,Minola2013,Hozoi2011} and Nd$_2$CuO$_4$ \cite{Kang2019} that possess the same square planar environment around the $d^9$ ion. Though, the $dd$ excitation peaks are broader than those of undoped cuprates and appear to deviate from a simple Gaussian or Lorentzian lineshape. These discrepancies may be due to the coupling to the particle-hole excitation continuum of the metallic state of Nd$_{1-x}$Sr$_x$NiO$_2$, and to the disorder associated with possible incomplete chemical reduction \cite{Lee2020,Goodge2020}. 

\begin{figure}
	\centering
	\includegraphics[width=\columnwidth]{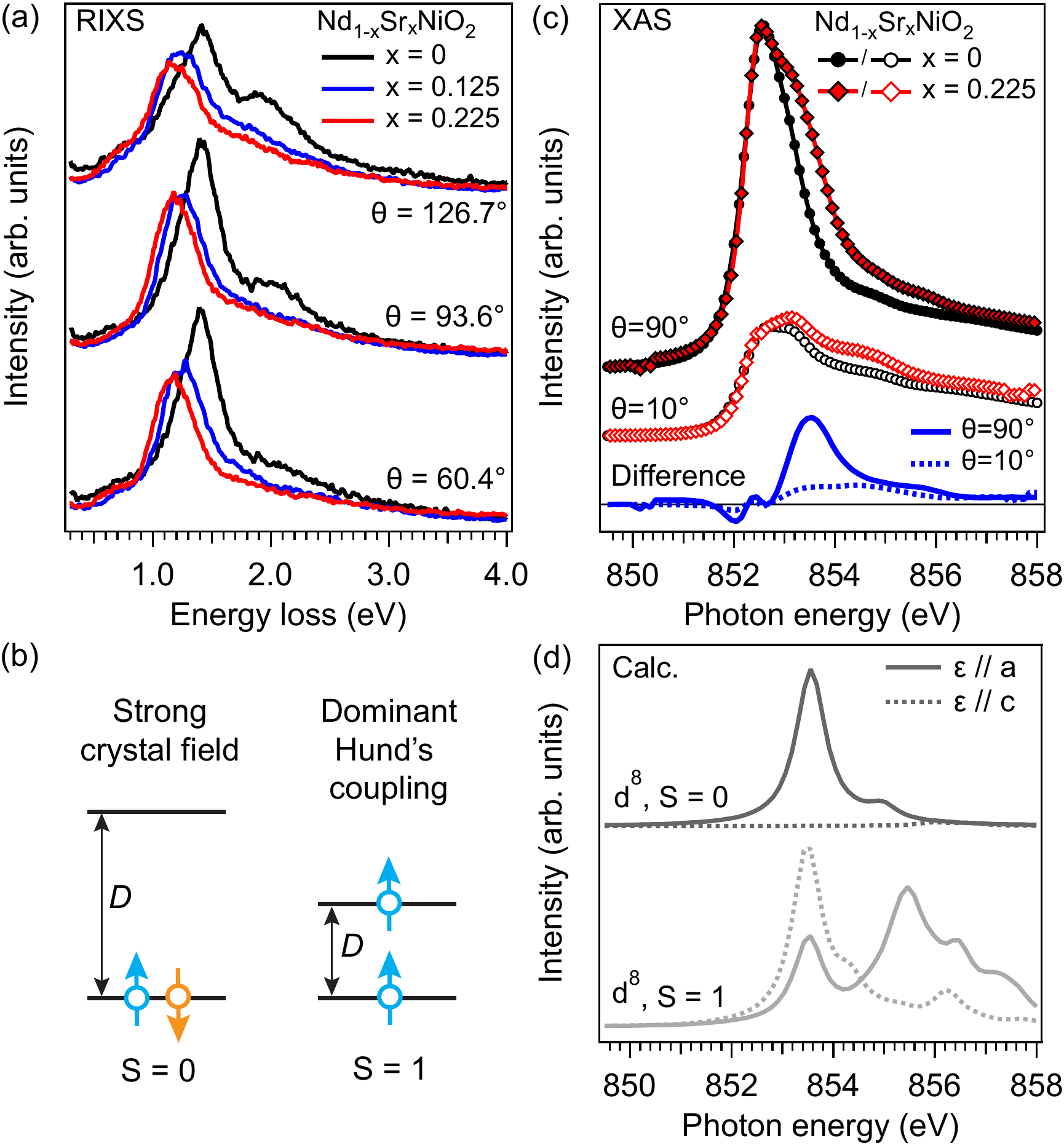}
	\caption{\label{fig4} (a) Doping dependence of the RIXS spectra of Nd$_{1-x}$Sr$_{x}$NiO$_2$ measured at various incidence angles with $\Theta = \SI{154}{\degree}$ and $\pi$ incident photon polarization. (b) Two-level diagrams showing the competition between Hund's coupling and crystal field splitting $D$. (c) XAS spectra of NdNiO$_2$ (black dots) and Nd$_{0.775}$Sr$_{0.225}$NiO$_2$ (red diamonds) measured at $\theta = \SI{90}{\degree}$ ($\bm{\epsilon} \parallel a$, filled markers) and $\theta = \SI{10}{\degree}$ ($\bm{\epsilon}$ almost parallel to $c$, empty marker). The spectra of the undoped compound are scaled and shifted to match the main peak of the doped sample. The difference between the spectra of Nd$_{0.775}$Sr$_{0.225}$NiO$_2$ and NdNiO$_2$ are plotted as blue lines. (d) Multiplet calculations of a $d^8$ site in spin-singlet ($S = 0$, dark grey) and spin-triplet ($S = 1$, light grey) configuration, respectively. Solid and dotted lines correspond to light polarization parallel to the $a$ and $c$ axes, respectively.}
\end{figure}

We now explore the evolution of orbital excitations upon doping. Figure~\ref{fig4}(a) shows the RIXS spectra of three samples with doping of $x$ = 0 (black), 0.125 (blue), and 0.225 (red) collected at three representative incidence angles. With increasing doping, the $dd$ excitations are broadened and, most importantly, softened by approximately 0.2 eV at $x = 0.225$. We note that the doping evolution of the $dd$ transitions is distinct from the case of hole-doped cuprates, where $dd$ excitations are broadened upon doping but their center of mass is little influenced \cite{Fumagalli2019}. This is because holes are doped into the O-derived charge-transfer band, which induces little effects in the energetics of the Cu $3d$ manifold. Conversely, in electron-doped cuprates the $dd$ excitations harden upon doping. This has been considered as a signature of electron doping into the Cu-derived upper Hubbard band, which effectively increases the energy difference between the Fermi level and other filled $3d$ orbitals \cite{Hepting2018}. Similarly, the softening of the $dd$ excitations in hole-doped infinite-layer niceklates indicates that doped carriers are introduced into the Ni $3d$ band corroborating with the shift of Fermi level toward the occupied $3d$ orbitals due to injection of holes.

What is the electronic state of the doped hole on Ni sites? Depending on the competition between crystal field splitting of the lowest energy $3d$ orbitals and Hund's rule coupling, doped holes on Ni sites (\textit{i.e.} a $d^8$ state) can arrange either in a spin singlet or triplet configuration, as exemplified in the two-level diagrams of Fig.~\ref{fig4}(b). A large Hund's exchange $J_H$ promotes the population of different orbitals in a high-spin state to minimize Coulomb repulsion (right diagram); instead, a strong crystal field $D$ favors a large orbital polarization where holes fill the first available state in a low-spin configuration (left diagram). To shed light on which scenario resembles the doped nickelates, we first extract the doping-induced change in the unoccupied states probed by Ni $L_3$ edge XAS. Figure~\ref{fig4}(c) displays XAS spectra of NdNiO$_2$ (black) and Nd$_{0.775}$Sr$_{0.225}$NiO$_2$ (red) measured with $\bm{\epsilon} \parallel a$ ($\theta = \SI{90}{\degree}$) and $\bm{\epsilon}$ mostly along the $c$ axis ($\theta = \SI{10}{\degree}$). Assuming that the main peak in both compounds originates from the $2p^63d^9$--$2p^53d^{10}$ transition, the spectra of NdNiO$_2$ are scaled and shifted by -0.05 eV and 0.05 eV for $\theta = \SI{90}{\degree}$ and $\theta = \SI{10}{\degree}$, respectively, to match the position and height of the first peak of Nd$_{0.775}$Sr$_{0.225}$NiO$_2$. The difference of the two spectra (blue lines) represents the additional contribution generated by doping; it mainly consists of a single peak that is strongly reduced when the polarization is mostly out of plane. Next, we calculate the multiplet spectrum of the $d^8$ ion in tetragonal crystal field in the two regimes, corresponding to spin triplet and spin singlet states. The parameters used, reported in the Supplemental Material \cite{SM}, are set to demonstrate the XAS gross features for the two respective scenarios which are robust against fine tuning. The multiplet spectra of the high- and low-spin states (Fig.~\ref{fig4}(d)) have distinct characteristics: the former is made of multiple peaks with similar weight over a wide energy range (light grey). When $\bm{\epsilon} \parallel c$, the first peak is enhanced, while the intensity of the high-energy features is reduced. Conversely, the multiplet spectrum of the spin-singlet configuration (dark grey) is dominated by a single peak that is strongly suppressed with out-of-plane polarization. Thus, the spin-singlet scenario where the doped hole occupies the $d_{x^2-y^2}$ orbital agrees with our data.

In conclusion, our results indicate that the doping evolution of the Ni $3d$ orbitals in infinite-layer nickelates is reminiscent of doping a single Hubbard band, namely, doped holes tend to reside in the $d_{x^2-y^2}$ orbital forming a $d^8$ spin singlet state. A low-spin configuration is highly uncommon for $d^8$ nickel compounds; however, it may become energetically favorable when the Ni ion is embedded in a low-symmetry environment, such as in planar complex K$_2$Ni-dithio-oxalate \cite{vanderLaan1988}, trilayer (La,Pr)$_4$Ni$_3$O$_8$ \cite{Zhang2017} and one-dimensional Sr$_2$Cu$_{0.9}$Ni$_{0.1}$O$_3$ \cite{Mandal2020}. Given the little change in O absorption spectra \cite{Goodge2020} and small electron pocket arisen from rare-earth $5d$ bands \cite{Lee2004}, it is likely that the low-energy electronic properties of Nd$_{1-x}$Sr$_{x}$NiO$_2$ are dominated by the Ni $3d_{x^2-y^2}$ orbitals. Note that the sign change of the Hall coefficient with temperature and doping \cite{Li2020,Zeng2020} has been interpreted as the contribution from multiple bands. However, a similar behavior is also observed in cuprates \cite{Tsukada2006,Armitage2010} where a single band description is well established. The complex dependence of the Hall coefficient may be related to electron correlations that influence the topology of the Fermi surface \cite{Wang2020}. Yet, the minor influence of other orbitals in the low-energy physics of infinite-layer nickelates remains an important open question. Nevertheless, our work provides a solid reference for future experimental and theoretical studies on the electronic structure of this new class of oxide superconductors. 

\begin{acknowledgments}
We are grateful to G.~A.~Sawatzky for insightful discussions. This work is supported by the U.S. Department of Energy (DOE), Office of Science, Basic Energy Sciences, Materials Sciences and Engineering Division, under contract DE-AC02-76SF00515. We acknowledge the Gordon and Betty Moore Foundation’s Emergent Phenomena in Quantum Systems Initiative through grant number GBMF4415 for synthesis equipment. We acknowledge Diamond Light Source for time on beamline I21-RIXS under Proposal NT25165 and MM25598. This research also used resources of the Advanced Light Source, a U.S. DOE Office of Science User Facility under contract no. DE-AC02-05CH11231.
\end{acknowledgments}

\bibliography{references}

\end{document}


\title{Supplemental Material\\ \quad \\Orbital and Spin Character of Doped Carriers in Infinite-Layer Nickelates}

\author{M.~Rossi}
\affiliation{Stanford Institute for Materials and Energy Sciences, SLAC National Accelerator Laboratory, 2575 Sand Hill Road, Menlo Park, California 94025, USA}

\author{H.~Lu}
\affiliation{Stanford Institute for Materials and Energy Sciences, SLAC National Accelerator Laboratory, 2575 Sand Hill Road, Menlo Park, California 94025, USA}
\affiliation{Department of Physics, Stanford University, Stanford, California 94305, USA}

\author{A.~Nag}
\affiliation{Diamond Light Source, Harwell Campus, Didcot OX11 0DE, United Kingdom}

\author{D.~Li}
\affiliation{Stanford Institute for Materials and Energy Sciences, SLAC National Accelerator Laboratory, 2575 Sand Hill Road, Menlo Park, California 94025, USA}
\affiliation{Department of Applied Physics, Stanford University, Stanford, California 94305, USA}

\author{M.~Osada}
\affiliation{Stanford Institute for Materials and Energy Sciences, SLAC National Accelerator Laboratory, 2575 Sand Hill Road, Menlo Park, California 94025, USA}
\affiliation{Department of Materials Science and Engineering, Stanford University, Stanford, California 94305, USA}

\author{K.~Lee}
\affiliation{Stanford Institute for Materials and Energy Sciences, SLAC National Accelerator Laboratory, 2575 Sand Hill Road, Menlo Park, California 94025, USA}
\affiliation{Department of Physics, Stanford University, Stanford, California 94305, USA}

\author{B.~Y.~Wang}
\affiliation{Stanford Institute for Materials and Energy Sciences, SLAC National Accelerator Laboratory, 2575 Sand Hill Road, Menlo Park, California 94025, USA}
\affiliation{Department of Physics, Stanford University, Stanford, California 94305, USA}

\author{S.~Agrestini}
\affiliation{Diamond Light Source, Harwell Campus, Didcot OX11 0DE, United Kingdom}

\author{M.~Garcia-Fernandez}
\affiliation{Diamond Light Source, Harwell Campus, Didcot OX11 0DE, United Kingdom}

\author{Y.-D.~Chuang}
\affiliation{Advanced Light Source, Lawrence Berkeley National Laboratory, 1 Cyclotron Road, MS 6-2100, Berkeley, California 94720, USA}

\author{Z.~X.~Shen}
\affiliation{Stanford Institute for Materials and Energy Sciences, SLAC National Accelerator Laboratory, 2575 Sand Hill Road, Menlo Park, California 94025, USA}
\affiliation{Department of Physics, Stanford University, Stanford, California 94305, USA}
\affiliation{Geballe Laboratory for Advanced Materials, Stanford University, Stanford, California 94305, USA}

\author{H.~Y.~Hwang}
\affiliation{Stanford Institute for Materials and Energy Sciences, SLAC National Accelerator Laboratory, 2575 Sand Hill Road, Menlo Park, California 94025, USA}
\affiliation{Department of Applied Physics, Stanford University, Stanford, California 94305, USA}
\affiliation{Geballe Laboratory for Advanced Materials, Stanford University, Stanford, California 94305, USA}

\author{B.~Moritz}
\affiliation{Stanford Institute for Materials and Energy Sciences, SLAC National Accelerator Laboratory, 2575 Sand Hill Road, Menlo Park, California 94025, USA}

\author{Ke-Jin~Zhou}
\affiliation{Diamond Light Source, Harwell Campus, Didcot OX11 0DE, United Kingdom}

\author{T.~P.~Devereaux}
\email[Corresponding author: ]{tpd@stanford.edu}
\affiliation{Stanford Institute for Materials and Energy Sciences, SLAC National Accelerator Laboratory, 2575 Sand Hill Road, Menlo Park, California 94025, USA}
\affiliation{Department of Materials Science and Engineering, Stanford University, Stanford, California 94305, USA}
\affiliation{Geballe Laboratory for Advanced Materials, Stanford University, Stanford, California 94305, USA}

\author{W.~S.~Lee}
\email[Corresponding author: ]{leews@stanford.edu}
\affiliation{Stanford Institute for Materials and Energy Sciences, SLAC National Accelerator Laboratory, 2575 Sand Hill Road, Menlo Park, California 94025, USA}


\maketitle

\subsection*{Oxygen \emph{K} edge XAS}

\begin{figure}
	\centering
	\includegraphics[width=0.75\columnwidth]{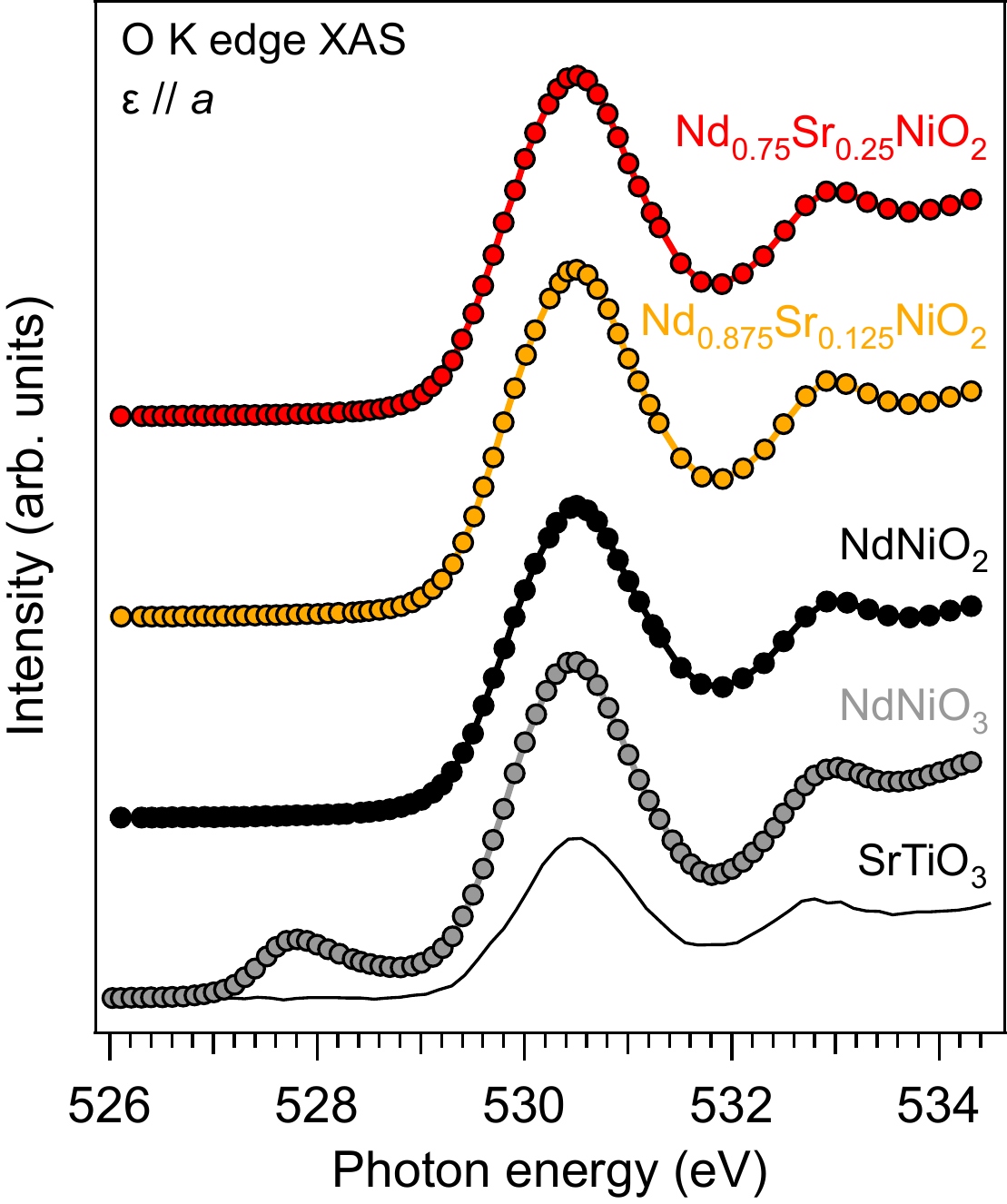}
	\caption{\label{figS1} Stack plot of total fluorescence yield O $K$ edge XAS spectra of Nd$_{1-x}$Sr$_x$NiO$_2$ (black: $x = 0$, orange: $x = 0.125$, red: $x = 0.25$) and NdNiO$_3$ (grey) measured with photon polarization parallel to the $a$ axis. The thin solid line is the XAS spectrum of SrTiO$_3$. Spectra are normalized to the intensity value at around 526 eV.}
\end{figure}

Oxygen $K$ edge XAS spectra were measured at beamline 8.0.1 of the Advanced Light Source (Berkeley, USA) using the qRIXS endstation. Spectra were collected at a temperature of 78 K. The XAS spectrum of SrTiO$_3$ is taken from Ref.~\onlinecite{Hepting2020}. The incident photon polarization was parallel to the sample $a$ axis. 

Figure~\ref{figS1} shows O $K$ edge XAS spectra of NdNiO$_2$ (black), Nd$_{0.875}$Sr$_{0.125}$NiO$_2$ (orange) and Nd$_{0.75}$Sr$_{0.25}$NiO$_2$ (red). The XAS signal is dominated by peaks from SrTiO$_3$ (thin black line), which has been used as substrate and capping layer. In particular, we do not detect the development of a pre-peak feature that is attributed to the doping-induced hole-population of the O-derived charge-transfer band \cite{Chen1991,Pellegrin1993}. To show the sensitivity of our XAS measurements to the signal from buried nickelate thin film, we plot in Fig.~\ref{figS1} the spectrum of a thin film of NdNiO$_3$ (grey dots) measured under the same conditions and with the same capping layer of five unit cells of SrTiO$3$ (we note that Nd$_{0.75}$Sr$_{0.25}$NiO$_2$ was grown with a thicker capping layer of 20 nm). A pre-edge feature at 527.8 eV is clearly visible in the XAS spectrum of NdNiO$_3$, due to charge transfer between the transition metal and ligand states \cite{Bisogni2016}.

Our O $K$ edge XAS measurements suggest that O-derived bands play a subdominant role in the low-energy physics of infinite-layer nickelates, consistent with spatially-resolved EELS measurements \cite{Goodge2020}. 

\subsection*{Neodymium \texorpdfstring{\emph{M}\textsubscript{5}}{\emph{M}5} edge XAS}

\begin{figure}
	\centering
	\includegraphics[width=0.85\columnwidth]{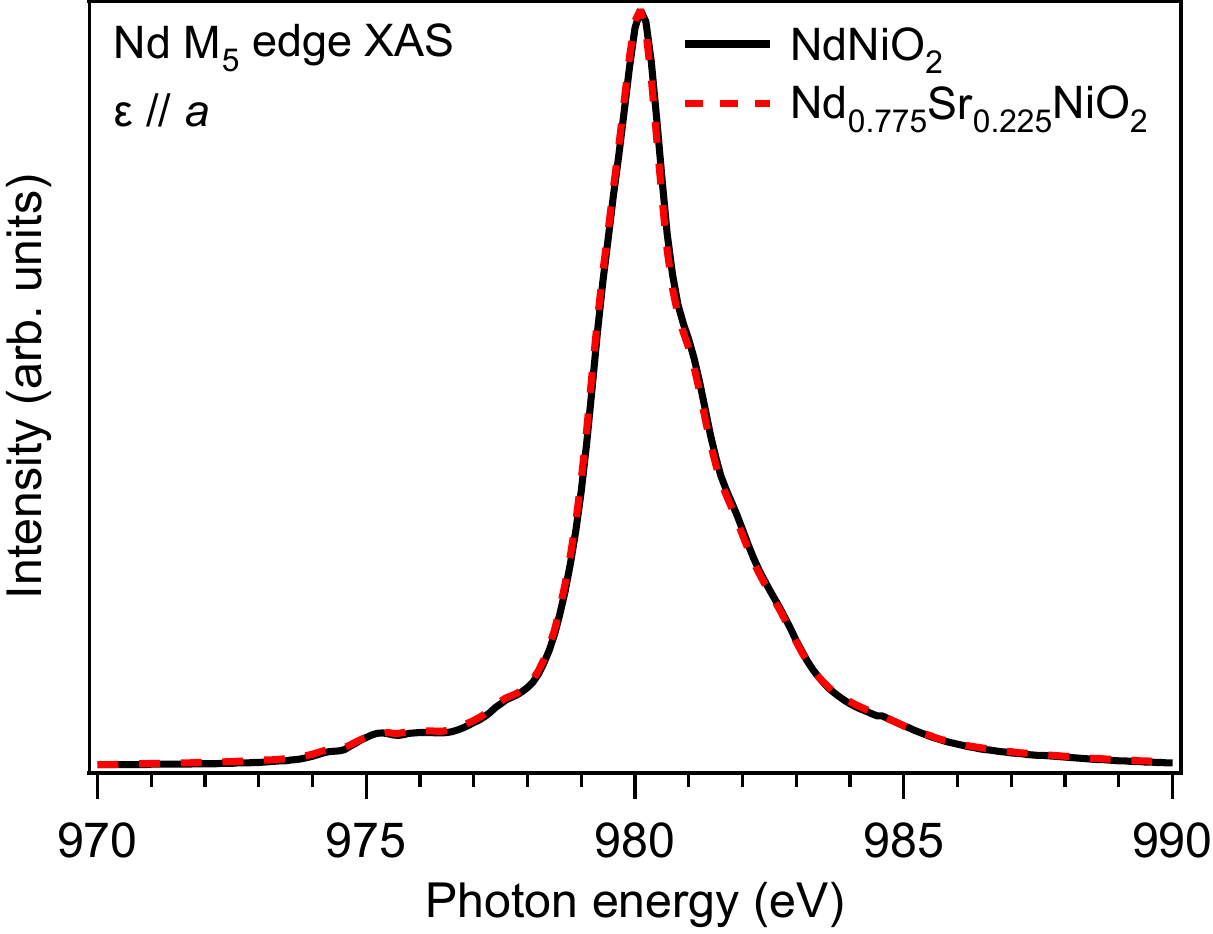}
	\caption{\label{figS2} Background-subtracted Nd $M_5$ edge XAS spectra of NdNiO$_2$ (black solid line) and Nd$_{0.775}$Sr$_{0.225}$NiO$_2$ (red dashed line) measured with photon polarization parallel to the $a$ axis. Spectra are collected in total electron yield mode and are normalized to the peak height.}
\end{figure}

Figure~\ref{figS2} shows the Nd $M_5$ edge XAS spectra of NdNiO$_2$ (black solid line) and Nd$_{0.775}$Sr$_{0.225}$NiO$_2$ (red dashed line). XAS spectra were collected at beamline I21 of the Diamond Light Source (United Kingdom). Measurements were taken at 20 K. The photon polarization vector was parallel to the sample $a$ axis. The spectra are almost identical and agree very well with the XAS spectrum of pure Nd \cite{Thole1985}, which has the same Nd$^{3+}$ valence configuration. The absence of doping-induced changes in the XAS spectra reveals that Nd $4f$ states are well localized and do not influence the relevant physics close to the Fermi level.

\subsection*{Single ion RIXS cross sections}

\begin{figure}
	\centering
	\includegraphics[width=0.8\columnwidth]{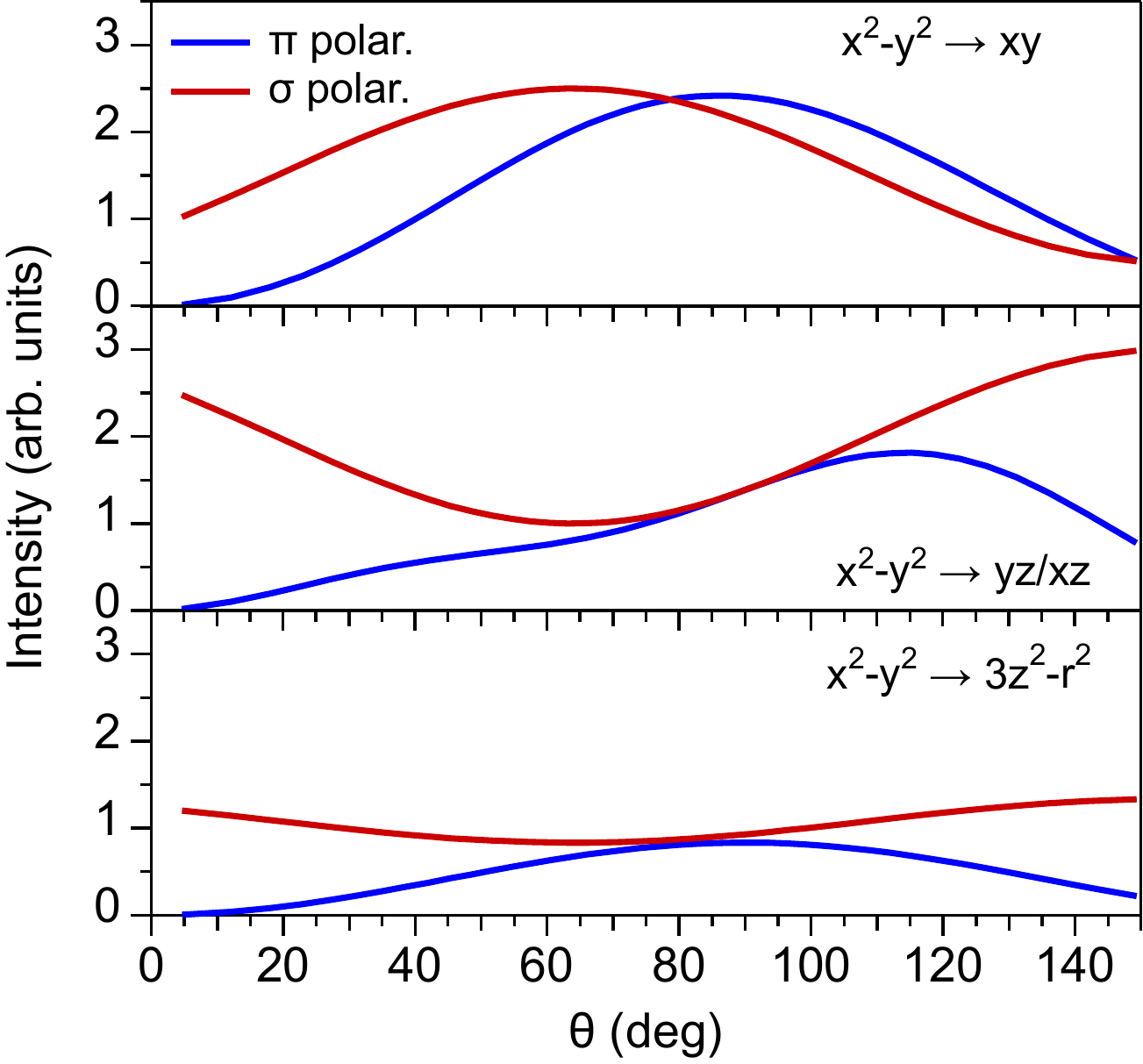}
	\caption{\label{figS3} Angular dependence of the $L_3$ edge RIXS cross section for transitions from the $d_{x^2-y^2}$ ground state to the final $d_{xy}$ (top panel), $d_{yz/xz}$ (middle) and $d_{3z^2-r^2}$ (bottom) orbitals. Blue and red lines correspond to RIXS intensities calculated for $\pi$ and $\sigma$ incident photon polarizations, respectively.}
\end{figure}

We consider the RIXS transition from the ground state $g$ with energy $E_g$ to the final state $f$ with energy $E_f$. The scattering amplitude $\mathcal{A}_{fg}$ is given by the Kramers-Heisenberg expression \cite{Ament2011}:
\begin{equation}
\label{eqS1}
    \mathcal{A}_{fg} = \sum_n \frac{\langle f | \mathcal{D}^{\prime\dagger} | n\rangle\langle n | \mathcal{D} | g \rangle}{E_g + \omega_\mathrm{in} - E_n +i \Gamma_n},
\end{equation}
where $n$ denotes the intermediate state with energy $E_n$ and lifetime broadening $\Gamma_n$; $\omega_\mathrm{in}$ is the incident photon energy; $\mathcal{D}$ and $\mathcal{D}^\prime$ are the dipole transition operators for photon absorption and emission, respectively. For a given edge, we consider that all intermediate states have approximately the same energy and lifetime, so that Eq.~\ref{eqS1} simplifies to:
\begin{equation}
\label{eqS2}
    \mathcal{A}_{fg} \approx \sum_n \langle f | \mathcal{D}^{\prime\dagger} | n\rangle\langle n | \mathcal{D} | g \rangle.
\end{equation}
We assume that the ground state is the atomic orbital $|x^2-y^2, 1/2\rangle$, where $1/2$ refers to the spin angular momentum. The final state can be one of the $d$ orbitals and includes both spin-conserving and spin-flip channels: $f = \{|xy, \pm 1/2\rangle, |yz/xz, \pm 1/2\rangle, |3z^2-r^2, \pm 1/2\rangle\}$. The direction of the spin quantization axis is irrelevant as long as the final state involves both spin channels. The dipole operator $\mathcal{D} = \bm{\epsilon}\cdot\mathbf{r}$ contains information on the electric field polarization vector $\bm{\epsilon}$ and thus on the scattering geometry. The polarization of the incoming photon is known, while the one of the scattered photon is not measured. Hence, the RIXS intensity is averaged over all outgoing polarizations that lie on a plane normal to the scattered photon wave vector. 

Figure~\ref{figS3} displays the $L_3$ edge RIXS cross sections for transitions into the final $d_{xy}$ (top panel), $d_{yz/xz}$ (middle) and $d_{3z^2-r^2}$ (bottom) orbitals as a function of incidence angle $\theta$ and for fixed scattering angle $\Theta = \SI{154}{\degree}$. Blue and red lines correspond to incident photon polarization parallel ($\pi$) and perpendicular ($\sigma$) to the scattering plane, respectively.

\subsection*{Multiplet calculations}

In the calculations, the atomic multiplet splitting is determined by the $3d$-$3d$ Slater-Condon parameters $F_{dd}^0$, $F_{dd}^2$ and $F_{dd}^4$, while the final state effects are included considering the $2p$ spin-orbit coupling (11.5 eV) and the interaction between the valence shell and the core hole, given by the $2p$-$3d$ Slater-Condon parameters $F_{pd}^0$, $F_{pd}^2$, $G_{pd}^1$, $G_{pd}^3$ \cite{deGroot1990}. We neglect hybridization between Ni and O orbitals in view of the large charge-transfer energy and weak weight of O bands in the ground state. The $D_{4h}$ crystal field splitting is obtained from the analysis of the RIXS $dd$ transitions of NdNiO$_2$. The energies of the $d$ orbitals relative to $d_{x^2-y^2}$ are: 1.39 eV for $d_{xy}$, 2.0 eV for $d_{yz/xz}$ and 2.7 eV for $d_{3z^2-r^2}$.

For the $d^8$ high-spin configuration, we used the Slater-Condon integrals for Ni$^{2+}$ from Ref.~\onlinecite{deGroot1990}, which correspond to $J_H = 1.0$ eV and Coulomb repulsion $U = 7.3$ eV. We then reduced these parameters in order to obtain a low-spin ground state, where both holes occupy the $d_{x^2-y^2}$ orbital. The set of parameters employed for the low-spin configuration were $J_H = 0.69$ eV and $U = 6$ eV, while the $2p$-$3d$ Slater-Condon parameters were $F_{pd}^0 = 3.0$ eV, $F_{pd}^2 = 3.09$ eV, $G_{pd}^1 = 2.315$ eV, and $G_{pd}^3 = 1.315$ eV.

The calculated spectra have been convoluted with a Lorentzian with full width at half maximum of 0.5 eV to take into account the $2p_{3/2}$ core hole lifetime \cite{Krause1979}.

\bibliography{references}